\documentclass[draft,grl]{agutex}

\pdfoutput=1
\usepackage{lineno}

\usepackage{amstext}
\usepackage{amsmath}
\usepackage{units}

\usepackage[pdftex,draft]{graphicx}
\setkeys{Gin}{draft=false}
\authorrunninghead{HEMINGWAY \& MATSUYAMA}

\titlerunninghead{ISOSTASY ON A SPHERE}

\authoraddr{Corresponding author: D. J. Hemingway,
Department of Earth and Planetary Science, University of
California, Berkeley, 94720, USA.
(djheming@berkeley.edu)}

\begin{document}

%
%

\title{Isostatic equilibrium in spherical coordinates and implications for crustal thickness on the Moon, Mars, Enceladus, and elsewhere}

%
%

%
%



\authors{Douglas J. Hemingway,\altaffilmark{1}
and Isamu Matsuyama,\altaffilmark{2}}

\altaffiltext{1}{Department of Earth and Planetary Science,
University of California, Berkeley, USA.}

\altaffiltext{2}{Lunar and Planetary Laboratory, University of Arizona,
Tucson, Arizona, USA.}

%
%



%
%


\begin{abstract}

Isostatic equilibrium is commonly defined as the state achieved
when there are no lateral gradients in hydrostatic pressure, and thus
no lateral flow, at depth within the lower viscosity mantle that underlies
a planetary body's outer crust. In a constant-gravity Cartesian framework,
this definition is equivalent to the requirement that columns of equal
width contain equal masses. Here we show, however, that this equivalence
breaks down when the spherical geometry of the problem is taken into
account. Imposing the ``equal masses'' requirement in a spherical
geometry, as is commonly done in the literature, leads to significant
lateral pressure gradients along internal equipotential surfaces,
and thus corresponds to a state of disequilibrium. Compared with the
``equal pressures'' model we present here, the ``equal masses''
model always overestimates the compensation depth\textemdash 
by $\sim$27\% in the case of the lunar highlands and by nearly a factor
of two in the case of Enceladus.

\end{abstract}

%
%

%

\begin{article}

%
%


\section{Introduction}

Rocky and icy bodies with radii larger than roughly $\unit[200]{km}$
typically have figures that are close to the expectation for hydrostatic
equilibrium (i.e., the surface conforms roughly to a gravitational
equipotential) because their interiors are weak enough that they behave
like fluids on geologic timescales. Because of high effective viscosities
in their cold exteriors, however, these bodies can 
maintain some non-hydrostatic topography, even on long
timescales. This non-hydrostatic topography may be
supported in part by bending and membrane stresses in the lithosphere
{[}e.g.,~\citealp{Turcotte1981}{]}, but
over long timescales, and especially when considering broad topographic
loads, or loads that formed at a time when the lithosphere was weak,
the rocks may yield until much of the support comes from
buoyancy\textemdash that is, the crustal material essentially floats
on the higher density, lower viscosity mantle material beneath it. This is the classic picture of isostatic equilibrium, first discussed
by Pratt and Airy in the 1850s, and is often invoked as a natural
mechanism by which gravity anomalies associated with topography can
be compensated {[}e.g.,~\citealp{Heiskanen1958,Watts2001}{]}. 
The two standard end-member models for isostatic
compensation are Airy, involving lateral variations in crustal thickness,
and Pratt, involving lateral variations in crustal density. 

The problem of modeling Airy-type isostatic compensation can be framed
as the need to compute the deflection of the interface between the
crust and the underlying higher density, lower viscosity material (we address Pratt-type compensation
in the Supporting Information, section~S2). Given the known surface
topography ($h_{t}$), the Airy-compensated basal topography ($h_{b}$)
can be computed as
\begin{linenomath*}
\begin{equation}
h_{b}=-h_{t}\frac{\rho_{c}}{\Delta\rho}\label{eq:Cartesian_isostasy}
\end{equation}
\end{linenomath*}
where $\rho_{c}$ is the density of the crustal material and $\Delta\rho$
is the density contrast at the crust/mantle interface. The negative
sign reflects the fact that the basal topography is inverted with
respect to the surface topography if both $h_{t}$ and $h_{b}$ are
taken as positive upward relief with respect to their respective reference
levels (i.e., the hypothetical equipotential surfaces to which the
density interfaces would conform if the layers were all inviscid).
This equation follows from requiring equal hydrostatic pressures at equal depths (or equivalently, requiring equal masses in columns of equal width), and ensures that, regardless
of the topography, there are no horizontal pressure gradients and
thus there is no lateral flow at depth within the fluid mantle (there
is also no vertical flow because vertical pressure gradients are balanced
by gravity). Hence\textemdash neglecting mantle dynamics and
the slow relaxation of the crust itself\textemdash we have
a state of equilibrium.

Equation~(\ref{eq:Cartesian_isostasy}) implicitly assumes a Cartesian
geometry and a uniform gravity field. However, for long wavelength loads or when the compensation depth is a substantial
fraction of the body's radius, it becomes necessary to take into account
the spherical geometry of the problem. In this case, the requirement of equal masses
in equal width columns leads to (section~\ref{subsec:Equal-Masses})
\begin{linenomath*}
\begin{equation}
h_{b}=-h_{t}\frac{\rho_{c}}{\Delta\rho}\left(\frac{R_{t}}{R_{b}}\right)^{2}\label{eq:equal_masses_isostasy}
\end{equation}
\end{linenomath*}
where $R_{t}$ and $R_{b}$ are the mean radii corresponding to the
top and bottom of the crust, respectively. This expression (or its
equivalent) is widely used in the literature {[}e.g.,~\citealp{Jeffreys1976a,Phillips1980,Hager1983,Lambeck1988,Wieczorek1997,Wieczorek2004,Hemingway2013,Mckinnon2015,Wieczorek2015b,Cadek2016,Cadek2017}{]}.
However, as we show in section~\ref{subsec:Equal-Pressures},
this is not equivalent to the requirement of equal pressures at equal
depths, which instead leads to
\begin{linenomath*}
\begin{equation}
h_{b}=-h_{t}\frac{\rho_{c}}{\Delta\rho}\left(\frac{g_{t}}{g_{b}}\right)\label{eq:equal_pressures_isostasy}
\end{equation}
\end{linenomath*}
where $g_{t}$ and $g_{b}$ are the mean gravitational accelerations
at the top and bottom of the crust, respectively. Although the distinction between ``equal masses" and ``equal pressures" isostasy has long been recognized {[}e.g.,~\citealp{Lambert1930,Heiskanen1958}{]}, it has widely been ignored because the effect is deemed negligible in the case of the Earth, where the crustal thickness is small compared to the radius. However, the difference between equations~(\ref{eq:equal_masses_isostasy}) and (\ref{eq:equal_pressures_isostasy})
becomes increasingly significant as the compensation depth becomes an increasingly
large fraction of the total radius, and can therefore be important
for bodies like the Moon, Mars, Ceres, Pluto and the outer solar system's
many mid-sized moons. 

Arguably, this basic picture of isostatic equilibrium suffers from some
internal inconsistencies in that, on one hand, it assumes that the
crust is stiff or viscous enough that the topography does not relax
away completely, while on the other hand assuming that the crust is
weak enough that it cannot support vertical shear stresses, meaning
that radial pressure gradients are the only available means of supporting
the topographic loads against gravity. 
Besides handling the spherical geometry properly, a fully self-consistent conception of the problem would have to account
for the internal stresses, the elastic and rheological
behaviors of the crust and mantle, the nature of the topographic loads
(i.e., where and when they were emplaced), and the system's 
time-varying response to those loads. 

Elastic stresses may prevent or at least slow the progression towards equilibrium, especially in
the case of relatively short-wavelength loads that deflect, but do
not readily break, the lithosphere. Accordingly, many authors construct analytical models based on thin elastic shell theory {[}e.g.,~\citealp{Kraus1967,Turcotte1981,Willemann1982,McGovern2002,Hemingway2013}{]}, wherein the loads are supported by a wavelength-dependent combination of bending stresses, membrane stresses, and buoyancy (in which the ``equal masses" versus ``equal pressures" distinction remains important). Still more sophisticated approaches exist as well. \citet{Beuthe2008}, for example, develops a more generalized analytical elastic shell model that allows for tangential loading and laterally variable elastic properties. Taking another approach, \citet{Belleguic2005} solves the elastic-gravitational problem numerically, accommodating the spherical geometry and the force balances in a self-consistent manner.

In the limit of a weak lithosphere (the isostatic limit), however, elastic stresses do not play such a significant role in supporting the topography. Some authors thus define isostatic equilibrium as the state of minimum deviatoric stresses within the lithosphere {[}e.g.,~\citealp{Dahlen1982,Beuthe2016a}{]}. This state is achieved in such models by splitting the crustal thickening (or thinning) into a suitable combination of surface and basal loads{\textemdash}in reality, the applied loads may have been entirely at the surface, entirely at the base, or some combination of the two; the combination that yields the state of minimum deviatoric stresses is merely intended to represent the final stress state after the lithosphere has finished failing or deforming in response to the applied loads. This approach aligns well with the basic concept of complete isostatic equilibrium in that it involves supporting the topography mainly by buoyancy, but with the additional advantage of maintaining internal consistency{\textemdash}deviatoric stresses do not go precisely to zero, and can thus keep the topography from relaxing away completely. Whereas implementation of this solution is far from straightforward {[}e.g.,~\citealp{Dahlen1982,Beuthe2016a} and references therein{]}, our simplified approach, in spite of its limitations, leads to a result that closely matches the minimum deviatoric stress result of \citet{Dahlen1982}.

One further consideration is the fact that relaxation may continue even after the initial gross isostatic adjustments have taken place. Provided that a topographic
load is broad, and that the underlying
layer is much weaker, the system will respond relatively rapidly at
first, on a timescale governed mainly by the viscosity of the underlying
weaker mantle, until reaching a quasi-static equilibrium in which
the lateral flow of that weak material is reduced to nearly zero. Relaxation does, however, continue after this point, and may not necessarily be negligible, especially when the base of the crust is relatively warm and ductile {[}e.g.,~\citealp{Zhong1997,Zhong2000,McKenzie2000,Zhong2002}{]}. Nevertheless, this latter stage of relaxation will usually be slow compared with the timescale for reaching isostatic equilibrium, and so we will often use the word ``equilibrium'' without qualification, even as we recognize the system may be continuing to evolve at some slow rate following the initial isostatic adjustment. We stress, however, that this is merely an assumption, and that caution should be used in cases where the materials are likely to relax more rapidly.

Notwithstanding the above complicating factors, the basic concept of isostatic equilibrium, in which topographic loads are supported entirely by buoyancy (i.e., without appeal to elastic stresses), has been widely and productively adopted as a useful approximation in Earth and planetary sciences. To the extent that such a simplified model remains desirable for analyses of planetary topography, it should at least be consistent with its core principle of avoiding lateral gradients in hydrostatic pressure at depth. This paper's modest goal is to show that, when accounting for the spherical geometry, the ``equal pressures" model, equation (\ref{eq:equal_pressures_isostasy}), provides a very good approximation that is consistent with this principle, while the commonly used ``equal masses" model, equation (\ref{eq:equal_masses_isostasy}), does not.

In section~\ref{sec:Analysis}, we show how we obtained equations
(\ref{eq:equal_masses_isostasy}) and (\ref{eq:equal_pressures_isostasy}),
and we compare the two in terms of the resulting internal pressure
anomalies. In section~\ref{sec:Implications}, we show how the two
different conceptions of isostasy affect spectral admittance and geoid-to-topography
ratio (GTR) models, addressing implications including crustal thickness
estimates for the specific examples of the lunar and Martian highlands,
as well as the ice shell thickness on Enceladus. Finally, we make
concluding remarks in section~\ref{sec:Conclusions}.

\section{Analysis\label{sec:Analysis}}

\subsection{Framework\label{subsec:Framework}}

Consider a body consisting of concentric layers, each having uniform density, and with the layer densities increasing monotonically inward. The shape of the $i^{th}$ layer can be expanded in spherical harmonics as
\begin{linenomath*}
\begin{equation}
H_{i}\left(\theta,\phi\right)=R_{i}+\sum_{l=1}^{\infty}\sum_{m=-l}^{l}H_{ilm}Y_{lm}\left(\theta,\phi\right)\label{eq:Hi}
\end{equation}
\end{linenomath*}
where $\theta$ and $\phi$ are the colatitude and longitude, respectively,
$Y_{lm}\left(\theta,\phi\right)$ are the spherical harmonic functions
for degree-$l$ and order-$m$ {[}e.g., \citealp{Wieczorek2015b}{]},
$R_{i}$ is the mean radius of the $i^{th}$ layer, and where the
coefficients $H_{ilm}$ describe the departure from spherical symmetry
for the $i^{th}$ layer. 

Each layer's shape is primarily a figure determined by hydrostatic equilibrium, but may include smaller additional non-hydrostatic topographic anomalies. Hence, we take the shape coefficients to be the sum of their hydrostatic and non-hydrostatic parts, $H_{ilm}=H_{ilm}^{\text{hyd}}+H_{ilm}^{\text{nh}}$. Since isostatic equilibrium concerns providing support for the departures from hydrostatic equilibrium, it is only the non-hydrostatic topographic anomalies, $H_{ilm}^{\text{nh}}$, that are involved in the isostatic equations. To a good approximation, the hydrostatic equilibrium figure can be described by a degree-2 spherical harmonic function. Hence, this complication generally does not apply to the topographic relief at degrees 3 and higher, where $H_{ilm}^{\text{hyd}}=0$. A possible exception is fast-rotating bodies, for which higher order hydrostatic terms may be non-negligible {[}\citealp{Rambaux2015}{]}.

We assume that the outermost shell (the ``crust") does not relax on the timescale relevant for achieving isostatic equilibrium, whereas we take the layer below the crust (the ``mantle") to be inviscid. Given the observed topographic relief at the surface, $H_{tlm}^{\text{nh}}$, we are concerned with finding the basal relief, $H_{blm}^{\text{nh}}$, required to deliver isostatic equilibrium. We consider the condition of isostatic equilibrium to be satisfied when there are no lateral variations in hydrostatic pressure along equipotential surfaces within the inviscid layer below the crust. The hydrostatic pressure at radial position $r$ is given by
\begin{linenomath*}
\begin{equation}
p\left(r,\theta,\phi\right)=\int_{r}^{\infty}\rho\left(r',\theta,\phi\right)g\left(r'\right)dr'\label{eq:pressure_at_depth}
\end{equation}
\end{linenomath*}
where $g\left(r\right)=GM\left(r\right)/r^{2}$ is the gravitational
acceleration at radius $r$, and where $M\left(r\right)$ is the enclosed
mass at radius $r$. Here, the small lateral variations in gravitational
acceleration are neglected. Although lateral variations in gravity can approach a few percent due to rotation and tidal forces, this simplification is justified on the grounds that the quantity of interest is often the ratio $g_t/g_b$, as in equation~(\ref{eq:equal_pressures_isostasy}) for example, and this ratio may be regarded as laterally constant.

A datum equipotential surface with mean radius $R_d$ can be approximated to first order as
\begin{linenomath*}
\begin{equation}
E_{d}\left(\theta,\phi\right)=R_{d}-\frac{\Delta{U}\left(R_d,\theta,\phi\right)}{g\left(R_{d}\right)}\label{eq:E_d}
\end{equation}
\end{linenomath*}
where ${g\left(R_{d}\right)}$ is the mean gravitational acceleration at $r=R_d$ and 
where $\Delta{U}\left(r,\theta,\phi\right)$ represents the lateral variations in the potential (section S1.3), given by
\begin{linenomath*}
\begin{equation}
\Delta{U}\left(r,\theta,\phi\right)=U^{\text{rot}}\left(r,\theta,\phi\right)+U^{\text{tid}}\left(r,\theta,\phi\right)+\sum_{l=1}^{\infty}\sum_{m=-l}^{l}U_{lm}\left(r\right)Y_{lm}\left(\theta,\phi\right)\label{eq:Delta_U}
\end{equation}
\end{linenomath*}
where $U^{\text{rot}}$ and $U^{\text{tid}}$ are the laterally varying rotational and (if applicable) tidal potentials, respectively, and where the coefficients $U_{lm}$ account for the gravitation associated with the topography and thus depend on the layer shapes and densities, and are given by
\begin{linenomath*}
\begin{equation}
U_{lm}\left(r\right)=-\frac{4\pi Gr}{2l+1}\sum_{i=1}^{N}\Delta\rho_{i}H_{ilm}\begin{cases}
\left(\frac{R_{i}}{r}\right)^{l+2} & r\geq R_{i}\\
\left(\frac{r}{R_{i}}\right)^{l-1} & r<R_{i}
\end{cases}\label{eq:Ulm}
\end{equation}
\end{linenomath*}
where $\Delta\rho_{i}$ is the density contrast between layer $i$ and the layer above it.

Below, we examine two distinct conceptions of the condition of Airy-type
isostasy in spherical coordinates: 1)~the requirement of equal masses
in columns (or cones) of equal solid angle; and 2)~the requirement of the absence
of lateral pressure gradients at depth, where pressure is assumed
to be hydrostatic. We use simplifying assumptions to obtain compact expressions for each case. We then evaluate these simple models by computing lateral pressure variations along the equipotential surface defined by (\ref{eq:E_d}). A good model should yield little or no lateral pressure gradients along this equipotential surface. For both models, we consider a two-layer body having a crust with density
$\rho_{\text{c}}$, and an underlying mantle with density $\rho_{m}$,
where $\rho_{m}>\rho_{c}$. For clarity and simplicity in the following derivations, we assume the body is not subjected to rotational or tidal deformation so that $H_{ilm}^{\text{hyd}}=0$. The top
and bottom of the crust have mean radii $R_{t}$ and $R_{b}$, respectively.
A portion of the body has some positive topographic anomaly at the
top of the crust ($h_{t}>0$) and a corresponding compensating isostatic
root (inverted topography) at the base of the crust ($h_{b}<0$) (Figure~S2a).
A reference datum is defined at an arbitrary internal radius $R_{d}<R_{b}+h_{b}$. 

\subsection{Equal Masses in Equal Columns\label{subsec:Equal-Masses}}

The mass above radius $r$, in any given column, taken as a narrow wedge, or cone, is given by
\begin{linenomath*}
\begin{equation}
M=\int_{r}^{\infty}\rho\left(r',\theta,\phi\right)r'^{2}\sin\theta d\theta d\phi dr'\label{eq:Mass}
\end{equation}
\end{linenomath*}
where $\theta$ and $\phi$ are colatitude and longitude, respectively. Equating the wedge mass in the absence of the topographic anomaly (left side of Figure~S2a) with the wedge mass in the presence of the topographic anomaly (right side of
Figure~S2a), yields
\begin{linenomath*}
\begin{equation*}
\Delta\rho\int_{R_{b}+h_{b}}^{R_{b}}r^{2}dr=\rho_{c}\int_{R_{t}}^{R_{t}+h_{t}}r^{2}dr
\end{equation*}
\end{linenomath*}
where $\Delta\rho=\rho_{m}-\rho_{c}$. After integrating, and some
manipulation, we obtain
\begin{linenomath*}
\begin{equation*}
h_{b}=-h_{t}\frac{\rho_{c}}{\Delta\rho}\left(\frac{R_{t}}{R_{b}}\right)^{2}\left(1+\frac{h_{t}}{R_{t}}+\frac{h_{t}^{2}}{3R_{t}^{2}}\right)\left(1+\frac{h_{b}}{R_{b}}+\frac{h_{b}^{2}}{3R_{b}^{2}}\right)^{-1}
\end{equation*}
\end{linenomath*}
If $\left|h_{t}\right|\ll R_{t}$ and $\left|h_{b}\right|\ll R_{b}$,
this expression reduces to equation~(\ref{eq:equal_masses_isostasy})
\begin{linenomath*}
\begin{equation*}
h_{b}\approx-h_{t}\frac{\rho_{c}}{\Delta\rho}\left(\frac{R_{t}}{R_{b}}\right)^{2}
\end{equation*}
\end{linenomath*}

\subsection{Equal Pressures at Depth\label{subsec:Equal-Pressures}}

Equating the hydrostatic pressure in the absence of the topographic anomaly (left side of Figure~S2a) with the hydrostatic pressure in the presence of the topographic anomaly (right
side of Figure~S2a), in both cases evaluated at $r=R_d$, we obtain
\begin{linenomath*}
\begin{equation*}
\Delta\rho\int_{R_{b}+h_{b}}^{R_{b}}g\left(r\right)dr=\rho_{c}\int_{R_{t}}^{R_{t}+h_{t}}g\left(r\right)dr
\end{equation*}
\end{linenomath*}
where again $\Delta\rho=\rho_{m}-\rho_{c}$. 

If $\left|h_{t}\right|\ll R_{t}$, then over the small radial distance
between $R_{t}$ and $R_{t}+h_{t}$, the integrand on the right hand
side has a nearly constant value of $g_{t}$,
the mean gravitational acceleration at $r=R_{t}$. Similarly, if $\left|h_{b}\right|\ll R_{b}$,
then on the left hand side, the integrand is always close to $g_{b}$,
the mean gravitational acceleration at $r=R_{b}$. Hence, if the relief
at the density interfaces is small, then it is a good approximation
to write
\begin{linenomath*}
\begin{equation*}
\Delta\rho g_{b}\int_{R_{b}+h_{b}}^{R_{b}}dr\approx\rho_{c}g_{t}\int_{R_{t}}^{R_{t}+h_{t}}dr
\end{equation*}
\end{linenomath*}
leading to equation~(\ref{eq:equal_pressures_isostasy})
\begin{linenomath*}
\begin{equation*}
h_{b}\approx-h_{t}\frac{\rho_{c}}{\Delta\rho}\left(\frac{g_{t}}{g_{b}}\right)
\end{equation*}
\end{linenomath*}

Because it is often more convenient to specify $\rho_{c}/\bar{\rho}$
(the ratio of the crustal density to the body's bulk density), it is useful to note that $g_{t}/g_{b}$ is given by (section~S1.5)
\begin{linenomath*}
\begin{equation}
\frac{g_{t}}{g_{b}}=\frac{\left(R_{b}/R_{t}\right)^{2}}{1+\left(\left(R_{b}/R_{t}\right)^{3}-1\right)\frac{\rho_{c}}{\bar{\rho}}}\label{eq:gt_over_gb}
\end{equation}
\end{linenomath*}

Note that the mass anomalies associated with the topographic anomaly and its compensating isostatic root will displace the datum equipotential surface slightly{\textemdash}an effect that is captured in (\ref{eq:E_d}), but which we have neglected in the derivation of equation~(\ref{eq:equal_pressures_isostasy}). If the radial displacement of this equipotential surface is $h_d$, the hydrostatic pressure at this depth (within the mantle) will be different by approximately $\rho_m g_d h_d$, where $g_d$ is the mean gravitational acceleration on this datum surface. 

For comparison, \citet{Turcotte1981} include the equivalent of this additional term (which they call $\rho_m g h_g$) in their pressure balance (their equation 3), though they neglect the radial variation in gravity and the fact that the shape of this equipotential surface will vary with depth (i.e., they evaluate $h_g$ only at the exterior surface, using their equation 25). In the limit of complete isostatic compensation, their $h_g$ goes to zero (substitute their eq.~28 into their eq.~25). Hence, in the isostatic limit, their equation 3 is identical to ours, except that we also account for the radial variation in gravity. 

In reality, due to the finite thickness of the crust, the displacement $h_d$ will not be precisely zero (it goes to zero for \citet{Turcotte1981} owing to some approximations they make to simplify their equation 25), but because we are concerned only with relatively small departures from hydrostatic equilibrium, $h_d$ is minuscule, and, as we show in the next section, in spite of our neglecting the $\rho_m g_d h_d$ term in the above derivation, our equation~(\ref{eq:equal_pressures_isostasy}) is nevertheless an excellent approximation when the goal is to make internal equipotential surfaces isobaric.

\subsection{Comparison}

In spite of the simplifications used to obtain equations (\ref{eq:equal_masses_isostasy})
and (\ref{eq:equal_pressures_isostasy}), it is clear that the two
results are not equivalent. To illustrate the difference, consider the case of a 2-layer body
(high viscosity crust, low viscosity mantle) that is initially spherically
symmetric (for simplicity, we again assume no tidal or rotational deforming potentials). We impose some topography
at the top of the crust, $H_{t}\left(\theta,\phi\right)$, and compute
the amplitude of the corresponding basal topography, $H_{b}\left(\theta,\phi\right)$,
using either (\ref{eq:Cartesian_isostasy}), (\ref{eq:equal_masses_isostasy}),
or (\ref{eq:equal_pressures_isostasy}). In each case, we then use
(\ref{eq:pressure_at_depth}) to compute the hydrostatic pressure at depth. 
Again, we are ultimately concerned with eliminating pressure gradients
along equipotential surfaces at depth, not just at a specific radial
position, so we compute internal pressure along the equipotential surface defined
by (\ref{eq:E_d}). 

\begin{figure}[p]
\begin{centering}
\includegraphics[width=11cm]{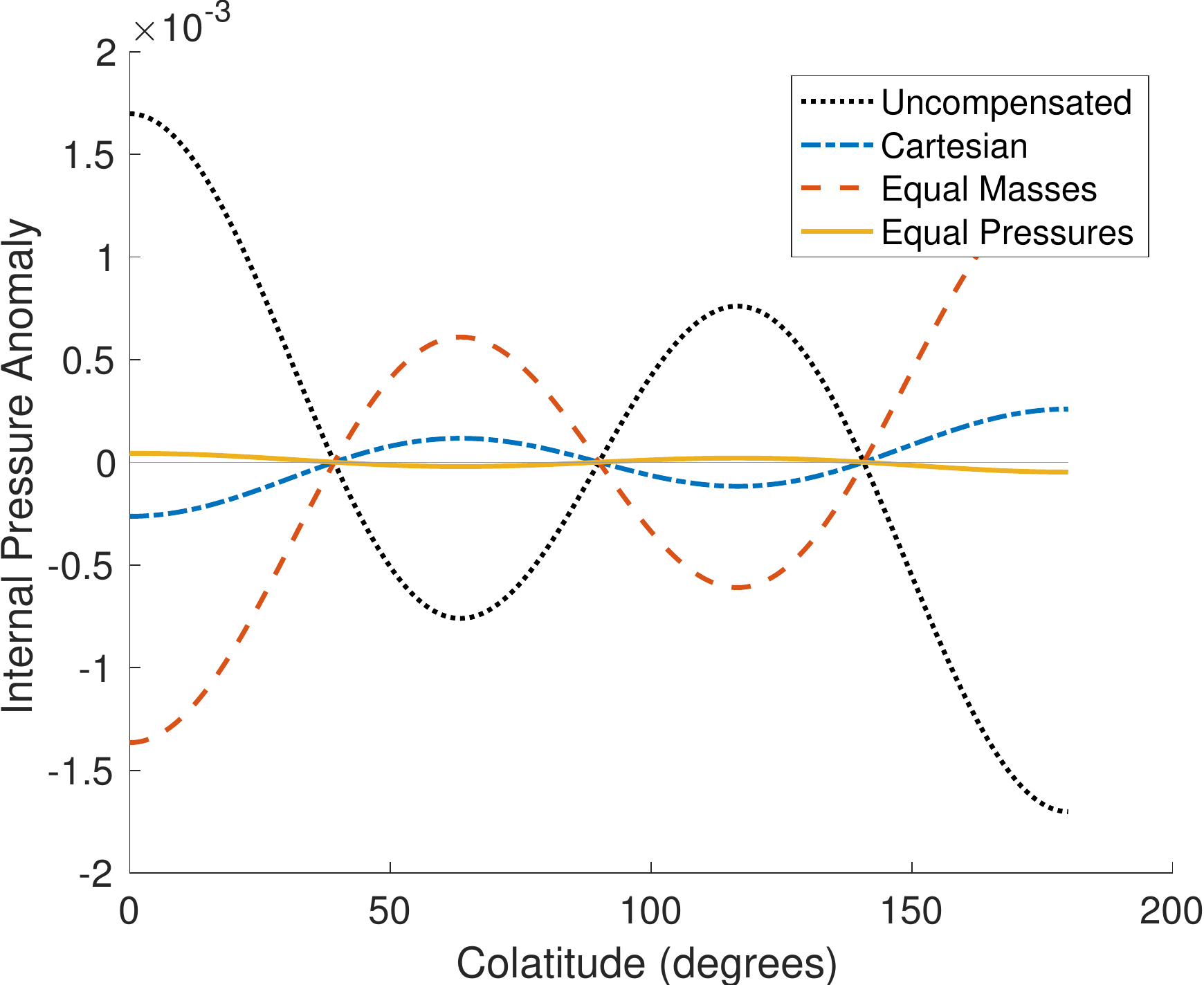}
\par\end{centering}
\caption{\label{fig:Internal-Pressure-Anomalies}Comparison of internal pressure
anomalies ($\delta p/\bar{p}$) for various basal topography solutions.
Zero pressure anomaly means zero lateral pressure gradients along
the equipotential surface $E_{d}\left(\theta,\phi\right)$.
The dotted black line illustrates the pressure anomaly resulting from
imposing the surface topography $H_{t30}$ without imposing any compensating
isostatic root (i.e., with $H_{b30}=0$). The colored lines illustrate
the pressure anomalies obtained when the isostatic root topography
($H_{b30}$) is computed via equations (\ref{eq:Cartesian_isostasy})
(Cartesian isostatic balance; dash-dotted blue), (\ref{eq:equal_masses_isostasy})
(equal mass in equal columns; dashed red), and (\ref{eq:equal_pressures_isostasy})
(equal pressures at depth; solid gold).}
\end{figure}

Figure~\ref{fig:Internal-Pressure-Anomalies} illustrates an example
in which the surface topography is described by a single non-zero
coefficient, $H_{t30}$, which is longitudinally symmetric, allowing
us to plot the internal pressure anomalies on an internal reference
equipotential surface as a function of colatitude
only. For reference, when the basal topography, $H_{blm}$, is zero, there
are of course significant lateral variations in pressure along the
equipotential surface, meaning we have a state of
disequilibrium (dotted black line in Figure~\ref{fig:Internal-Pressure-Anomalies}).
When the topography is compensated according to equation~(\ref{eq:Cartesian_isostasy}),
the pressure anomalies are reduced, but not eliminated (dash-dotted
blue line). When the topography is compensated according to equation~(\ref{eq:equal_masses_isostasy}),
the internal pressures change substantially, but large lateral pressure
gradients remain, and so we still have a state of disequilibrium (dashed
red line). When the topography is compensated according to equation~(\ref{eq:equal_pressures_isostasy}),
on the other hand, the lateral pressure gradients nearly vanish (solid
gold line), as expected if the assumptions made in section~\ref{subsec:Equal-Pressures}
are reasonable.
Hence, only equation~(\ref{eq:equal_pressures_isostasy}) describes
a condition that is close to equilibrium. In this example, we arbitrarily
set $\rho_{c}=\unit[1000]{kg/m^{3}}$, $\rho_{m}=\unit[3000]{kg/m^{3}}$,
$R_{t}=\unit[100]{km}$, $R_{b}=\unit[80]{km}$, such that $\rho_{c}/\bar{\rho}\approx0.49$,
$R_{d}=\unit[50]{km}$, and we impose a topographic anomaly with amplitude
$H_{t30}=\unit[200]{m}$, 1\% of the mean crustal thickness. The fundamental
conclusions are not, however, sensitive to these choices: compared
with equation~(\ref{eq:equal_pressures_isostasy}), equation~(\ref{eq:equal_masses_isostasy})
always gives rise to larger pressure anomalies.

When compensation depths are shallow, $g_{t}\approx g_{b}$ and $R_{t}\approx R_{b}$,
so that equations (\ref{eq:equal_masses_isostasy}) and (\ref{eq:equal_pressures_isostasy})
both reduce to the usual Cartesian form of the isostatic balance.
However, when compensation depths
become non-negligible fractions of the body's total radius, equations
(\ref{eq:Cartesian_isostasy}), (\ref{eq:equal_masses_isostasy}),
and (\ref{eq:equal_pressures_isostasy}) begin to diverge. When the
crustal density is less than $\sim70\%$ of the body's bulk density,
then $g_{t}<g_{b}$ (section~S1.5, Figure~S1),
meaning that equation~(\ref{eq:Cartesian_isostasy}) generally overestimates
the amplitude of the basal topography. When the crustal density is
more than $\sim70\%$ of the body's bulk density (as is likely the
case for Mars, for example), $g_{t}$ may be larger than $g_{b}$,
and so equation~(\ref{eq:Cartesian_isostasy}) could underestimate
the amplitude of the basal topography. However, of the three equations,
(\ref{eq:equal_masses_isostasy}) always yields the largest (most
overestimated) isostatic roots because $R_{t}>R_{b}$ and because,
assuming density does not increase with radius, $\bar{\rho}\leq\bar{\rho}_{b}$
(section~S1.5). 

\section{Implications\label{sec:Implications}}

\subsection{Spectral Admittance\label{subsec:Spectral-Admittance}}

In combined studies of gravity and topography, it is common to use
the spectral admittance as a means of characterizing the degree or
depth of compensation {[}e.g., \citealp{Wieczorek2015b}{]}. The mass
associated with any surface topography (represented using spherical
harmonic expansion coefficients, $H_{tlm}$) produces a corresponding
gravity anomaly. However, if the topography is compensated isostatically\textemdash that
is, if there is some compensating basal topography ($H_{blm}$)\textemdash the
gravity anomaly can be reduced. 

Using equation~(S13), we can compute the surface gravity anomaly caused by the topography at the top and bottom
of the crust, yielding
\begin{linenomath*}
\begin{equation}
g_{lm}=\frac{l+1}{2l+1}4\pi G\left(\rho_{c}H_{tlm}+\Delta\rho H_{blm}\left(\frac{R_{b}}{R_{t}}\right)^{l+2}\right)\label{eq:Delta_glm}
\end{equation}
\end{linenomath*}
where again, $\rho_{c}$ is the density of the crust, $\Delta\rho$
is the density contrast at the crust/mantle interface, 
and where we have neglected any contributions that may arise from
asymmetries on deeper density interfaces. 

Taking the degree-$l$ admittance, $Z_{l}$, to be the ratio of gravitational
acceleration ($g_{lm}$) to topography ($H_{tlm}$), and assuming complete
Airy compensation, with the basal topography ($H_{blm}$) computed
via the ``equal masses'' model, equation~(\ref{eq:equal_masses_isostasy}),
we have
\begin{linenomath*}
\begin{equation}
Z_{l}=\frac{l+1}{2l+1}4\pi G\rho_{c}\left(1-\left(\frac{R_{b}}{R_{t}}\right)^{l}\right)\label{eq:Z_from_equal_masses}
\end{equation}
\end{linenomath*}

Equation~(\ref{eq:Z_from_equal_masses}) is commonly used to generate
a model admittance spectrum under the assumption of complete Airy
compensation.
Comparison of the model admittance with the observed admittance, along
with an assumption about the crustal density then allows
for an estimate of the compensation depth, $d=R_{t}-R_{b}$. 

However, when we instead compute the basal topography using the ``equal
pressures'' equation~(\ref{eq:equal_pressures_isostasy}), we obtain
\begin{linenomath*}
\begin{equation}
Z_{l}=\frac{l+1}{2l+1}4\pi G\rho_{c}\left(1-\left(\frac{g_{t}}{g_{b}}\right)\left(\frac{R_{b}}{R_{t}}\right)^{l+2}\right)\label{eq:Z_from_equal_pressures}
\end{equation}
\end{linenomath*}
where again $g_{t}/g_{b}$ is given by equation~(\ref{eq:gt_over_gb}).

\begin{figure}[p]
\begin{centering}
\includegraphics[width=16cm]{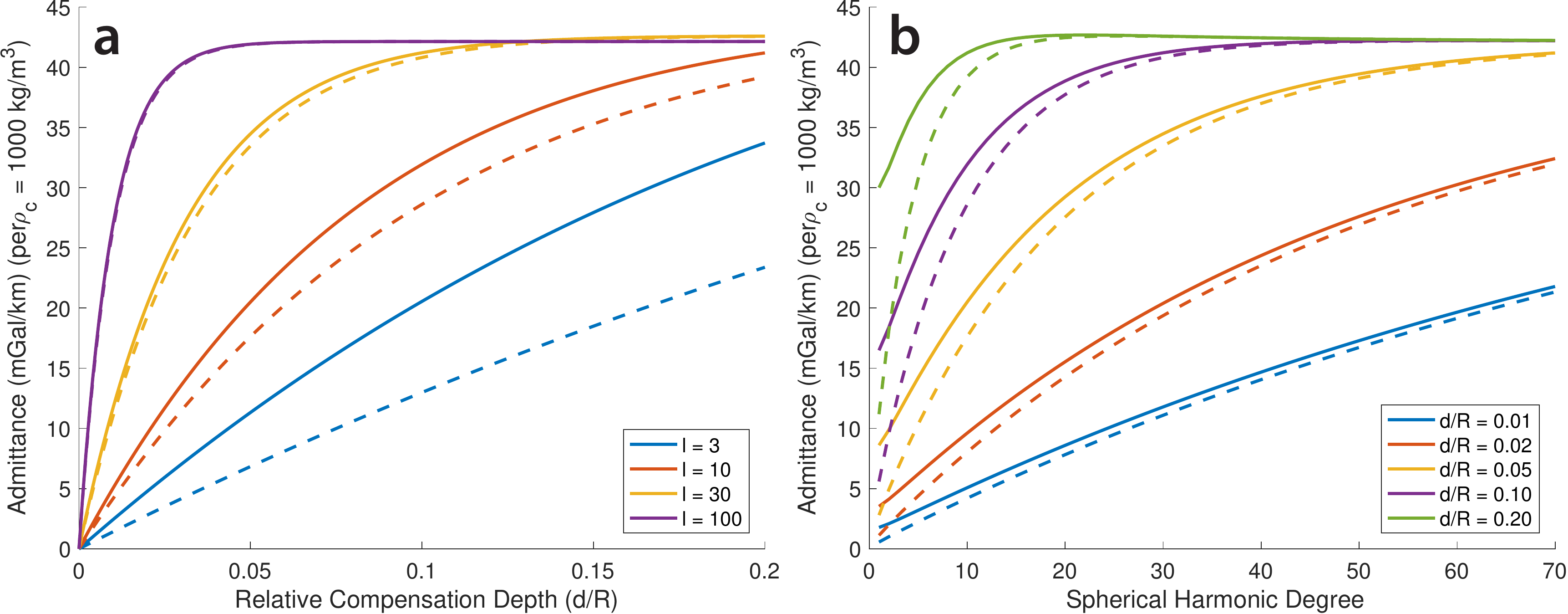}
\par\end{centering}
\caption{\label{fig:Z_vs_d_and_l}Admittance assuming Airy compensation. (a)
Admittance as a function of relative compensation depth ($d/R$) for
various example spherical harmonic degrees. (b) Spectral admittance
for various examples of relative compensation depths. Dashed lines
show admittance as computed via (\ref{eq:Z_from_equal_masses}), which
assumes equal masses in equal columns. Solid lines show admittance
as computed via (\ref{eq:Z_from_equal_pressures}), which eliminates
lateral pressure gradients at depth. The ``equal masses'' conception
of isostasy always leads to underestimating the admittance, especially
at low spherical harmonic degrees (long wavelengths). In both panels,
admittance is normalized to an assumed crustal density of $\unit[1000]{kg/m^{3}}$
(i.e., if the crustal density is $\unit[2000]{kg/m^{3}}$, all admittance
values double). Equation~(\ref{eq:Z_from_equal_pressures}) also
depends weakly on the internal density structure, which is here arbitrarily
defined by $\rho_{c}/\bar{\rho}=0.6$.}
\end{figure}

Compared with equation~(\ref{eq:Z_from_equal_pressures}), equation~(\ref{eq:Z_from_equal_masses})
will always lead to an overestimate of the compensation depth. That
is, at any given spherical harmonic degree, using equation~(\ref{eq:Z_from_equal_pressures})
yields the same admittance with a smaller compensation depth (Figure~\ref{fig:Z_vs_d_and_l}a).
Equivalently, for any given compensation depth, the model admittance
spectrum computed via equation~(\ref{eq:Z_from_equal_pressures})
is larger than that obtained via equation~(\ref{eq:Z_from_equal_masses})
(Figure~\ref{fig:Z_vs_d_and_l}b). The discrepancy is always greatest
at low spherical harmonic degrees (e.g., focusing on degree 3, and
assuming that $\rho_{c}/\bar{\rho}=0.6$, would yield a compensation
depth estimate that is roughly $\sim50\%$ too large) and vanishes
in the short wavelength limit (e.g., the compensation depth overestimate
reduces to $<5$\% for $l>50$). 

For clarity and simplicity, we have not included the finite amplitude
(or terrain) correction {[}e.g., \citealt{Wieczorek1998}{]} in the
above admittance equations. When the topographic relief is a non-negligible
fraction of the body's radius, it may be important to include this
effect, which will in general lead to larger admittances. However,
the point of this paper is not so much to advocate the use of equation~(\ref{eq:Z_from_equal_pressures})
in the admittance calculation, but rather, more fundamentally, to
advocate the use of equation~(\ref{eq:equal_pressures_isostasy})
in computing the basal topography.

It is worth emphasizing that the degree-2 admittance is complicated
by the effects of rotational and possibly tidal deformation. A meaningful
admittance calculation for degree-2 requires first removing the tidal/rotational
effects from both the gravity and topography signals. Only the remaining,
non-hydrostatic, signals should then be used in the admittance calculation.
Unfortunately, determination of the hydrostatic components of the
degree-2 gravity and topography signals requires knowledge of the
body's interior structure, which may not be readily available. In
such cases, the easiest option would be to simply exclude the degree-2
terms in the admittance analysis. Alternatively, one might appeal
to self-consistency arguments to constrain the internal structure
and admittance simultaneously {[}e.g., \citealt{Iess2014}{]}.

\subsection{Geoid-to-Topography Ratio (GTR)\label{subsec:Geoid-to-Topography-Ratio-(GTR)}}

A closely related concept is the geoid-to-topography ratio (GTR),
which has been used to estimate regional crustal thicknesses in situations
where local isostasy can be reasonably expected [e.g., \citealt{Wieczorek1997,Wieczorek2004}].
\citet{Wieczorek1997} showed that the GTR is primarily a function
of crustal thickness and can be computed from a compensation model
according to 
\begin{linenomath*}
\begin{equation}
\text{GTR}=R_{t}\sum_{l=l_{\text{min}}}^{l_{\text{max}}}W_{l}Q_{l}\label{eq:GTR}
\end{equation}
\end{linenomath*}
where $W_{l}$ is a weighting coefficient for degree-$l$, and $Q_{l}$
is a transfer function relating the degree-$l$ gravitational potential
and topography coefficients
\begin{linenomath*}
\begin{equation}
Q_{l}=\frac{C_{lm}}{H_{lm}}\label{eq:Q_of_l}
\end{equation}
\end{linenomath*}

The weighting coefficients reflect the fact that the geoid is most
strongly affected by the longest wavelengths (lowest spherical harmonic
degrees) and are constructed based on the topographic power spectrum,
$S_{hh}\left(l\right)=\sum_{m=-l}^{l}H_{lm}^{2}$, according to 
\begin{linenomath*}
\begin{equation}
W_{l}=S_{hh}\left(l\right)/\sum_{i=l_{\text{min}}}^{l_{\text{max}}}S_{hh}\left(i\right)\label{eq:W_of_l}
\end{equation}
\end{linenomath*}
 \citep{Wieczorek2015b}. $Q_{l}$ may be regarded as another expression
for the spectral admittance ($Z_{l}$), except that it employs dimensionless
gravitational potential coefficients rather than acceleration, and
so we denote it here with a distinct symbol (also in accord with \citet{Wieczorek2015b}). 

Neglecting the effects of topography on boundaries other than the
surface and the crust/mantle interface, we can use equation~(S12)
to rewrite (\ref{eq:Q_of_l}) as
\begin{linenomath*}
\begin{equation}
Q_{l}=\frac{3}{2l+1}\left(\frac{\rho_{c}}{R_{t}\bar{\rho}}\right)\left(1+\frac{\Delta\rho H_{blm}}{\rho_{c}H_{tlm}}\left(\frac{R_{b}}{R_{t}}\right)^{l+2}\right)\label{eq:Q_of_l_general}
\end{equation}
\end{linenomath*}

Assuming complete Airy compensation, with the basal topography ($H_{blm}$)
computed via the ``equal masses'' equation~(\ref{eq:equal_masses_isostasy}),
we then have
\begin{linenomath*}
\begin{equation}
\text{GTR}=\sum_{l=l_{\text{min}}}^{l_{\text{max}}}W_{l}\left(\frac{3}{2l+1}\right)\left(\frac{\rho_{c}}{\bar{\rho}}\right)\left(1-\left(\frac{R_{b}}{R_{t}}\right)^{l}\right)\label{eq:GTR_em}
\end{equation}
\end{linenomath*}

If we instead compute the basal topography using the ``equal pressures''
equation~(\ref{eq:equal_pressures_isostasy}), we obtain
\begin{linenomath*}
\begin{equation}
\text{GTR}=\sum_{l=l_{\text{min}}}^{l_{\text{max}}}W_{l}\left(\frac{3}{2l+1}\right)\left(\frac{\rho_{c}}{\bar{\rho}}\right)\left(1-\left(\frac{g_{t}}{g_{b}}\right)\left(\frac{R_{b}}{R_{t}}\right)^{l+2}\right)\label{eq:GTR_ep}
\end{equation}
\end{linenomath*}

For reference, the linear dipole moment approximation \citep{Ockendon1977,Haxby1978}
can be written
\begin{linenomath*}
\begin{equation}
\text{GTR}=\left(\frac{3}{2}\right)\left(\frac{\rho_{c}}{\bar{\rho}}\right)\left(1-\frac{R_{b}}{R_{t}}\right)\label{eq:GTR_dm}
\end{equation}
\end{linenomath*}

\begin{figure}[p]
\begin{centering}
\includegraphics[width=10cm]{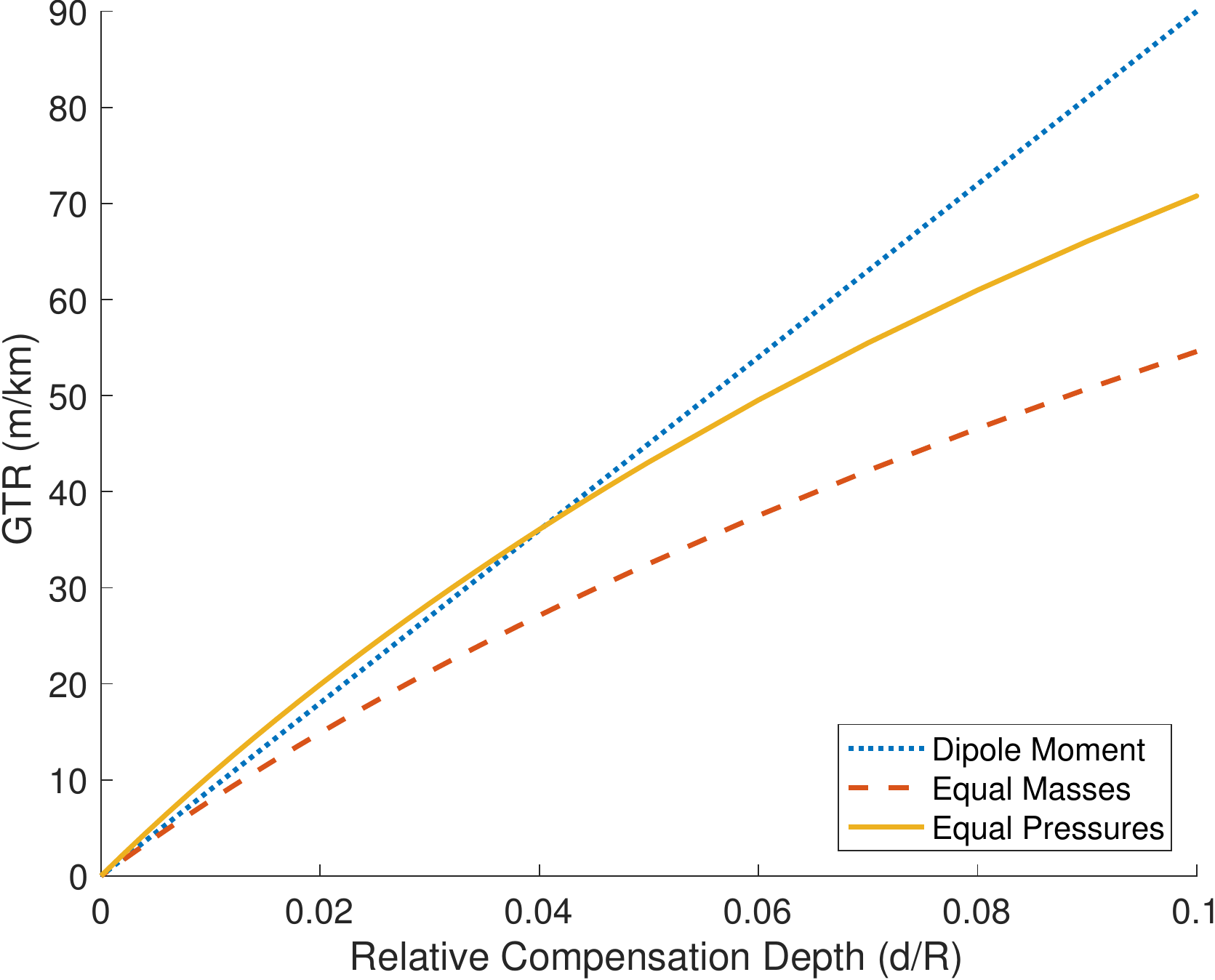}
\par\end{centering}
\caption{\label{fig:GTR_vs_d}Geoid-to-Topography Ratio (GTR) as a function
of relative compensation depth ($d/R$). Dotted blue line shows GTR
computed via (\ref{eq:GTR_dm}), using the linear dipole moment approximation.
Dashed red line shows GTR computed via (\ref{eq:GTR_em}), which assumes
equal masses in equal columns. Solid gold line shows GTR computed
via (\ref{eq:GTR_ep}), which avoids lateral pressure gradients at
depth. The internal density structure is again arbitrarily defined
by $\rho_{c}/\bar{\rho}=0.6$. The sum in (\ref{eq:GTR}) is taken
from $l_{\text{min}}=3$ to $l_{\text{max}}=70$. The weighting coefficients
are obtained from (\ref{eq:W_of_l}) by assuming a synthetic power
spectrum defined by $S_{hh}=Al^{-1.5}$, where $A$ is an arbitrary
constant. (\textit{cf}.~Figure~3a in \citet{Wieczorek1997} and
Figure~1 in \citet{Wieczorek2004}.)}
\end{figure}

Each model thus suggests a different relationship between the GTR
and the compensation depth (Figure~\ref{fig:GTR_vs_d}). For shallow
compensation depths (i.e., less than $\sim4\%$ of the body's radius
assuming $\rho_{c}/\bar{\rho}=0.6$), the ``equal pressures'' conception
of isostasy and the linear dipole moment approximation give similar
results. For deeper compensation depths, the dipole moment approach
begins to overestimate the GTR. In all cases, the ``equal masses''
approach underestimates the GTR, and therefore leads to an overestimate
of the compensation depth (Figure~\ref{fig:GTR_vs_d}).

\subsection{Application to the Moon, Mars, and Icy Satellites\label{subsec:Application}}

Here we consider a few realistic examples to illustrate how crustal
thickness estimates differ when one adopts the ``equal pressures''
rather than the ``equal masses'' model.
Note that the ``equal pressures''-based crustal thickness values
discussed in this section should not be taken as definitive new estimates.
There are many subtleties to the interpretation of gravity and topography
data that we have ignored here. The tools discussed in sections~\ref{subsec:Spectral-Admittance}
and \ref{subsec:Geoid-to-Topography-Ratio-(GTR)} will comprise only
one component of any meaningful analysis of planetary crusts. \citet{Wieczorek2004},
for instance, provide a comprehensive analysis that incorporates geochemical
and mechanical equilibrium considerations to complement their GTR
analysis. An updated estimate of the Martian highlands crustal thickness
would require careful consideration of a wide range of relevant factors
and an exploration of the permissible parameter space. Here, we wish
only to illustrate, using a few specific examples, the importance
of adjusting the admittance and GTR components of the analysis to
incorporate the ``equal pressures'' isostatic equilibrium model
rather than the ``equal masses'' model.

For the case of the nearside lunar highlands, \citet{Wieczorek1997}
obtained geoid-to-topography ratios (GTRs) of roughly $\unit[14-34]{m/km}$.
Taking the case of a single layer crust (\citet{Wieczorek1997}
also considered dual-layer crusts), with a density of $\unit[2900]{kg/m^{3}}$
($\rho_{c}/\bar{\rho}\approx0.87$), this yields a crustal thickness
estimate of roughly $\unit[22-61]{km}$ when the topography is assumed
to be in isostatic equilibrium in the ``equal masses'' sense. Adopting
the ``equal pressures'' model instead leads to crustal thickness
estimates of $\unit[18-48]{km}$, suggesting that the ``equal masses''
model overestimates the crustal thickness by up to $\sim27\%$ in
this case (section~S3.1, Figure~S6a). For the Martian highlands, \citet{Wieczorek2004} obtained
GTRs of roughly $\unit[13-19]{m/km}$, corresponding to crustal thicknesses
of roughly $\unit[48-73]{km}$, assuming a crustal density of $\unit[2900]{kg/m^{3}}$
($\rho_{c}/\bar{\rho}\approx0.74$) and adopting the ``equal masses''
approach. The ``equal pressures'' model instead leads to crustal
thicknesses of roughly $\unit[44-66]{km}$, not as dramatically different
as in the case of the lunar highlands, but still indicating that the
``equal masses'' model overestimates the crustal thickness by $\sim10\%$
in the case of the Martian highlands (section~S3.1, Figure~S6b). 

For icy bodies, the ice shell's density can be a considerably smaller
fraction of the bulk density, leading to smaller $g_{t}/g_{b}$ ratios
and therefore even more pronounced differences between the ``equal
masses'' and ``equal pressures'' isostasy models (Figure~S3).
In the case of Europa, for example, a crustal density of $\unit[930]{kg/m^{3}}$
corresponds to $\rho_{c}/\bar{\rho}\approx0.31$, leading the crustal
thickness estimates to differ by a factor of roughly two at the lowest
spherical harmonic degrees. For Encleadus ($\rho_{c}/\bar{\rho}\approx0.58$,
assuming $\unit[\rho_{c}=930]{kg/m^{3}}$), where the degree-2 and
-3 gravity terms have been measured based on a series of \textit{Cassini}
flybys, \citet{Iess2014} were able to obtain a degree-3 admittance
of $\unit[14.0\pm2.8]{mGal/km}$, which allows for a crustal thickness estimate of
$\unit[30\pm6]{km}$, adopting the ``equal masses'' model.
Adopting the ``equal pressures'' model instead leads to a remarkably
different estimate of just $\unit[17\pm4]{km}$ (section~S3.2,
Figure~S7).

\section{Conclusions\label{sec:Conclusions}}

To the extent that isostatic equilibrium is a useful model
for the state of mature planetary crusts, where broad topographic
loads are supported mainly by buoyancy, it should be taken to mean a
state in which hydrostatic (or lithostatic) pressures are equal along
equipotential surfaces within the relatively low viscosity mantle.
However, it is common in the literature to define isostatic equilibrium
as the requirement that columns of equal width contain equal masses.
Whereas these two definitions would be equivalent in a Cartesian framework,
we have shown here that they are not equivalent in a spherical geometry
(section~\ref{sec:Analysis}). We have demonstrated that adopting
the ``equal masses'' model leads to lateral pressure gradients that
can be nearly as large (though opposite in sign) as if there were
no isostatic compensation at all (Figure~\ref{fig:Internal-Pressure-Anomalies}).
We further showed that the ``equal masses'' model leads to an overestimate
of either the compensating basal topography in the case of Airy compensation
(section~\ref{sec:Analysis}), or the compensating lateral crustal
density variations in the case of Pratt compensation (section~S2). 

In combined studies of gravity and topography, using an ``equal masses''
model leads to an overestimate of the compensation depth (Figures~\ref{fig:Z_vs_d_and_l}
and S4). The discrepancy is always most significant at the lowest
spherical harmonic degrees (longest wavelengths) and increases as the crustal density becomes a smaller fraction of the
body's bulk density. As examples, we showed that, in the case of the lunar and
Martian highlands, the ``equal masses'' model could overestimate
the crustal thicknesses by $\sim27\%$ and $\sim10\%$, respectively.
For the case of Enceladus, where the compensation depth may be on
the order of $10\%$ of the radius and where the ice shell density is
roughly $58\%$ of the bulk density, the ``equal masses'' model
may overestimate the shell thickness by nearly a factor of two. In the case of asymmetric loads (odd harmonics), we additionally note that the ``equal masses'' and ``equal pressures'' models will lead to distinct center of mass-center of figure offsets, a factor that could be important for smaller bodies.

Whereas, for the sake of clarity, we have focused here on the end-member
case of complete isostatic equilibrium (purely buoyant support), the
distinction between ``equal masses'' and ``equal pressures'' remains
important for models in which the topography is supported by a combination
of both buoyancy and elastic flexure\textemdash a topic that is beyond the 
scope of this work. While we acknowledge the limitations of the very concept
of isostatic equilibrium (see Introduction), our
goal here is merely to ensure that isostasy models at least correspond to
what they are intended to mean\textemdash no lateral flow at depth
when topographic loads are supported entirely by buoyancy. That is,
in order to be consistent with the basic principle of isostasy, we
must be sure to use the ``equal pressures'' model presented here
and not the ``equal masses'' model. Beyond this simple picture,
a fully self-consistent model of a planetary crust and its topography
requires consideration of its loading history (i.e., where and when
the loads were emplaced), the state of internal stresses (and failures)
through time, and the potentially time-varying rheology of the relevant
materials, within both the crust and the underlying mantle. Such models
could be highly valuable, but only where sufficient clues are available
to meaningfully constrain these many factors. In the absence of such
information, the condition of isostatic equilibrium, as we have presented
it here, is likely to remain a useful model, at least as a reference
end member case.


%
%
%
%
%
%
%

\begin{acknowledgments}
This work was initially motivated by a discussion with Bill McKinnon and also benefited from exchanges with Bruce Buffet, Anton Ermakov, Roger Fu, Michael Manga, Tushar Mittal, Francis Nimmo, Gabriel Tobie, and especially Mikael Beuthe. We thank Mark Wieczorek and Dave Stevenson for constructive reviews that improved the manuscript. All data are publicly available, as described in the text. Financial support was provided by the Miller Institute for Basic Research in Science at the University of California Berkeley, and the NASA Gravity Recovery and Interior Laboratory Guest Scientist Program.
\end{acknowledgments}

%
%
%
%
%
%
%
%
%


\nocite{Hubbard1984,Hemingway2016,Iess2010,Nimmo2011a}

\bibliographystyle{agu08}
\bibliography{library}
\end{article}
%
%
%
%
%
%
%
%


\end{document}